\documentclass[a4paper]{article}
\usepackage{epsfig}
\usepackage{graphicx}  

\date{\today}
\newcommand{\bmat}{\left(\begin{array}}
\newcommand{\emat}{\end{array}\right)}
\newcommand{\be}{\begin{equation}}
\newcommand{\ee}{\end{equation}}
\newcommand{\bea}{\begin{eqnarray}}
\newcommand{\eea}{\end{eqnarray}}

\def\eps{\epsilon}
\def\al{\alpha}

\def\n{\nu}

\def\lam{\lambda}
\def\th{\theta}

\begin{document}
\title{\Large\bf $\left(Z_2\right)^3$ Symmetry of the Tripartite Model}
\author
{ \it \bf  E. I. Lashin$^{1,2,3}$\thanks{elashin@ictp.it}, E.
Malkawi$^{4}$ \thanks{emalkawi@uaeu.ac.ae}, S. Nasri$^{4}$ \thanks{snasri@uaeu.ac.ae} and N.
Chamoun$^{5}$\thanks{nidal.chamoun@hiast.edu.sy} ,
\\ \small$^1$ The Abdus Salam ICTP, P.O. Box 586, 34100 Trieste, Italy. \\
\small$^2$ Ain Shams University, Faculty of Science, Cairo 11566,
Egypt.\\ \small$^3$ Department of physics and Astronomy, College
of Science, King Saud University, Riyadh, Saudi Arabia,\\
\small$^4$  Department of Physics, UAE University, P.O.Box 17551,
Al-Ain, United Arab Emirates, \\
\small$^5$  Physics Department, HIAST, P.O.Box 31983, Damascus,
Syria.  }

\maketitle
\begin{center}
\small{\bf Abstract}\\[3mm]
\end{center}
We derive in a simple way the $\left(Z_2\right)^3$ symmetry which characterizes
uniquely the phenomenologically successful tripartite form leading
to the tribimaximal mixing in the neutrino mass matrix. We impose
this symmetry in a setup including the charged leptons and find
that it can accommodate all the possible patterns of lepton
masses in the framework of type-I and type-II seesaw mechanisms. We also discuss the
possibility of generating enough baryon asymmetry
through lepton-lepton asymmetry.
\\\\
{\bf Keywords}: Neutrino Physics; Flavor Symmetry; Matter-anti-matter.
\\
{\bf PACS numbers}: 14.60.Pq; 11.30.Hv; 98.80.Cq
\begin{minipage}[h]{14.0cm}
\end{minipage}
\vskip 0.3cm \hrule \vskip 0.5cm

\section{Introduction}
The present data from neutrino oscillations determine
approximately the $3 \times 3$ neutrino mixing matrix with the two
mass-squared differences \cite{SK,SNO,CHOOZ}.  In the standard
model (SM) of particle interactions, there are 3 lepton families.
The charged-lepton mass matrix linking left-handed to their
right-handed counterparts is arbitrary, but can always be
diagonalized by a bi-unitary transformation:
\begin{equation}
{M}_l = U^l_L \pmatrix{m_e & 0 & 0 \cr 0 & m_\mu & 0 \cr 0 & 0 &
m_\tau} (U^l_R)^\dagger.
\end{equation}
Likewise, we can diagonalize the neutrino mass matrix either by a
bi-unitary transformation if the neutrinos are of Dirac-type:
\begin{equation}
{M}^D_\nu = U^\nu_L \pmatrix{m_1 & 0 & 0 \cr 0 & m_2 & 0 \cr 0 & 0
& m_3} (U^\nu_R)^\dagger,
\end{equation}
or by just one unitary transformation if it is symmetric, which is
the case when the neutrinos are of Majorana-type:
\begin{equation}
{M}^M_\nu = U^\nu_L \pmatrix{m_1 & 0 & 0 \cr 0 & m_2 & 0 \cr 0 & 0
& m_3} (U^\nu_L)^T.
\end{equation}
 The
observed neutrino mixing matrix comes from the mismatch between
$U^l_L$ and $U^\nu_L$ in that
\begin{eqnarray}
U_{l\nu} = (U^l_L)^\dagger U^\nu_L \simeq \pmatrix{0.83 & 0.56 &
<0.2 \cr -0.39 & 0.59 & 0.71 \cr -0.39 & 0.59 & -0.71} .
\end{eqnarray}
We see that this mixing matrix is approximately equal to a
specific pattern $V_0$ named tribimaximal by Harrison, Perkins,
and Scott (HPS) \cite{HPS}:
\begin{eqnarray}
U_{l\nu} \simeq V_0 &\equiv&
\pmatrix{\sqrt{2/3} & 1/\sqrt{3} & 0 \cr -1/\sqrt{6} & 1/\sqrt{3}
& 1/\sqrt{2} \cr 1/\sqrt{6} & -1/\sqrt{3} & +1/\sqrt{2}}.
\end{eqnarray}

One might ask whether or not the HPS tribimaximal form $V_0$, with its rich phenomenology \cite{tbm}, 
results from a symmetry. If we work in the flavor basis where $M_l M_l^\dagger$
is diagonal, thus
$U_L^l = {\bf 1}$ is a unity matrix, and  assume the neutrinos
are of Majorana-type, then the flavor mixing matrix is simplified
to $V_0 \simeq U_{l\nu}= U_L^\nu$, and so, with $M^{\mbox{diag}}_\n = \mbox{Diag}\left( m_1, m_2, m_3\right)$, 
we have
\bea M_\n &=& V_0 \cdot M^{\mbox{diag}}_\n \cdot V_0^T.
\label{star}
\eea 
A special form
of $M_{\nu}$ in this basis was proposed by Ma in \cite{matripar}
which leads to the tribimaximal mixing. This ``tripartite'' form is
\begin{equation}
\label{trisum}
{M}_{\nu} = { M}_A + { M}_B + { M}_C,
\end{equation}
where
\begin{eqnarray}
{M}_A = A \pmatrix {1 & 0 & 0 \cr 0 & 1 & 0 \cr 0 & 0 & 1}, ~~ {
M}_B = B \pmatrix {-1 & 0 & 0 \cr 0 & 0 & 1 \cr 0 & 1 & 0}, ~~ {
M}_C = C \pmatrix {1 & 1 & -1 \cr 1 & 1 & -1 \cr -1 & -1 & 1},
\label{formz}
\end{eqnarray}
with neutrino eigen masses:
\begin{eqnarray}
\label{neigenm}
m_1=A-B, \; m_2 =
A-B + 3C, \; m_3 = A+B.
\end{eqnarray}

It was shown in \cite{matripar} that $M_A+M_B$ respect a
$Z_3\times Z_2$ symmetry, but the $Z_3$ is broken under $M_C$, and
the whole mass matrix $M_{\nu
}$ is $Z_2$ symmetric. One can show \cite{lmc}
that $M_A+M_B$ has a $U(1)$ underlying
symmetry which is broken by $M_C$, however a residual $\left(Z_2\right)^3$ symmetry
is invariably left unbroken
and this latter symmetry characterizes uniquely the tripartite
form in Eq.~(\ref{trisum}).

In fact, it was pointed out in \cite{lavsy} that any
neutrino mass matrix possessed a $\left(Z_2\right)^3$ symmetry (reduced to $Z_2\times Z_2$ when restricted to 
the ``special'' symmetries preserving the ``orientation'' of the flavor basis), which meant that
this ``general'' kind of symmetry was devoid of any physical significance since it is just a mere
consequence of the diagonalizability of $M_\n$. However, one should notice that the specific form of
this $\left(Z_2\right)^3$ depends
on the form of the neutrino mass matrix. Hence, when we consider some restricted forms of neutrino mass matrix
 (the tripartite model in our case) then the associated symmetry will clearly have a physical significance
 related to the form under study. Moreover, the symmetry of the HPS tribimaximal neutrino mixing was also
 examined in \cite{king-luhn} and was shown to originate from a pattern $M_\n$ which can be cast
 in the tripartite form. In this paper, we rederive these results in a simple way by imposing
 the form invariance idea \cite{maform} and examining its implications on the diagonalized neutrino
 mass matrix $M^{\mbox{diag}}_\n$ as the analysis in the latter case is very simple. Nonetheless, any symmetry
 defined in the basis
$(\nu_e,\nu_\mu,\nu_\tau)$ is automatically applicable to
$(e,\mu,\tau)$ in the complete Lagrangian, and thus we show, in addition, that one can
implement this symmetry on the lepton sector coupled to additional
scalar fields with suitably chosen Yukawa couplings and get in a natural manner the charged-lepton 
mass hierarchy. Furthermore, if one assumes
the canonical seesaw mechanism with the Majorana neutrino mass
matrix,
\begin{equation}
\label{Majmass}
{M}_\nu = -{ M}^D_\nu { M}_R^{-1} ({ M}^D_\nu)^T,
\end{equation}
where $M_R$ is the heavy Majorana mass matrix
for the right neutrinos, then we show that one can accommodate the different possible mass
patterns.

The plan of the paper is as follows. In section 2 we find a realization of the
$\left(Z_2\right)^3$ symmetry underlying the tripartite, and hence the HPS tribimaximal, model.
 In section 3 we introduce extra scalar fields and derive the charged-lepton mass matrix. In section 4,
we infer within the framework of type-I seesaw mechanism the neutrino mass matrix, whereas in section 5 
we discuss another scenario for the neutrino mass matrix in the framework of type-II seesaw
mechanism. In both scenarios we discuss also the possibility of explaining Baryon asymmetry 
using lepton asymmetry. We end up by summarizing our results in section 6.

\section{{\large \bf The underlying symmetry of the HPS tri-bimaximal pattern}}
The approach of form invariance states that the neutrino mass matrix is invariant when expressed in the
flavor basis and another basis related to the former by a specific
unitary transformation $S$:
\begin{equation}
\label{form}
S^T {M_\n}
S = {M_\n}.
\end{equation}
In order to find the most general symmetry $S$ that imposes
the form invariance property on a given $M_\n$, we see that
this invariance is equivalent, using equation (\ref{star}), to
\begin{equation}
\label{formdiag}
U^T M^{{\mbox{diag}}}_\n U = M^{{\mbox{diag}}}_\n,
\end{equation}
where $U$ is a unitary matrix related to $S$ by
\bea S&=&V_0^*\cdot U\cdot V_0^T. \eea
This means that the diagonalized mass matrix is itself form invariant under
$U$, and any $U$ symmetry for the
diagonalized form can appear as an $S$ symmetry in the flavor
basis. However, equation (\ref{formdiag}) implies
\bea
\label{commutator}
U^*\,M^{\mbox{diag}}_\n &=& M^{\mbox{diag}}_\n\, U,\eea
and the experimental data, implying three distinct masses ($m_1, m_2, m_3$), would force $U$ to
be of the form:
\bea U=\mbox{Diag}(\pm 1, \pm 1, \pm 1)\label{U}.\eea

The ``geometrical'' interpretation of the symmetry group is now very clear in the diagonalized basis,
in that we have a group $U$ formed of eight elements $U=\{\pm I,\pm I_i\}$ where $I_i$ represents
the $i$ reflection $\left[i=x,y,z : {\it {e.g.}}\; I_x=\mbox{Diag}(-1,+1,+1)\right]$. Since the inverse of 
any element in $U$
is itself and since $U$ is generated by the three reflections (note that $-I = I_x I_y I_z$ and, say, that
$-I_z = I_x I_y$) we
may write:
\bea U &=& \langle I_x,I_y,I_z\rangle \cong \left(Z_2\right)^3.\eea
The determinant-function from $U$ to the multiplicative group $\left(\{-1,+1\}\right)$ is a group morphism
whose kernel forms a subgroup $U_0$ of $U$ consisting of the unitary matrices satisfying the form invariance
property  and whose determinant is equal to $1$, and is generated by two elements, say:
\bea U_0 &=& \langle -I_x,-I_y\rangle \cong \left(Z_2\right)^2.\eea

One can thus find a realization of $\left( Z_2\right)^3$ (or of $Z_2^2$) for any pattern in the flavor basis
characterized by the mixing matrix $V$ by simply writing
 \bea S&=&V^* \cdot \mbox{Diag}(\pm 1, \pm 1, \pm 1) \cdot V^T. \eea
 Thus, the three generators of the $\left( Z_2\right)^3$ characteristic of the HPS tribimaximal mixing are
\bea
\label{Z2cubesymmetry1}
S_1 &=V_0^*\,\cdot I_x \cdot \,V_0^T=& \frac{1}{3}\pmatrix {-1 & 2 & -2 \cr 2 & 2 & 1 \cr -2 & 1
& 2}, \\  S_2 &=V_0^*\cdot I_y \cdot V_0^T=& \frac{1}{3}\pmatrix {1 & -2 & 2 \cr -2 & 1 & 2 \cr 2 & 2
& 1}, \\ S_3 &=V_0^*\cdot I_z \cdot V_0^T=& \pmatrix {1 & 0 & 0 \cr 0 & 0 & -1 \cr 0 & -1
& 0}. \label{Z2cubesymmetry3}\eea
We have discovered here, in a simple perspective, the $\left( Z_2\right)^3$ symmetry characteristic of the
tribimaximal mixing. Moreover, we could check that the tripartite form ---which, as noted above, is equivalent
to the tribimaximal pattern--- can be determined uniquely by the subgroup $S_0=\langle -S_1, -S_2\rangle$, or the
whole group $S \cong \left( Z_2\right)^3$ (since $U$ appears quadratically in equation \ref{formdiag}):
\bea 
\label{equivalence}
\left( S_i^T.M.S_i=M \right) & \Leftrightarrow & \left( \exists A,B,C: M=\left ( \matrix{ A - B + C
& C & -C \cr C & A+C & B-C \cr -C & B-C & A+C \cr} \right )
\right),
\eea 
where $i=1,2$ or $i=1,2,3$.
Additionally, we note here that this proposed structure can be altered minimally in order to accommodate 
deviation from the tripartite form. In fact, had the experimental data given a maximal atmospheric mixing 
angle $\th_y=\frac{\pi}{4}$ and a solar mixing angle $\th_x=\al$ different from the tribimaximal 
value $\th_{x_0}=\arctan{\left(\frac{1}{\sqrt{2}}\right)}$, then the mixing matrix $V$ would
be
\bea \label{xgeneral} V_\al =  R_{23}(\theta_y=\frac{\pi}{4}) \otimes
R_{12}(\theta_x = \al)  & = & \left ( \matrix{ c_\al      & s_\al & 0 \cr
-\frac{s_\al}{\sqrt{2}} & \frac{c_\al}{\sqrt{2}}   &
\frac{1}{\sqrt{2}} \cr \frac{s_\al}{\sqrt{2}} &
-\frac{c_\al}{\sqrt{2}}  & \frac{1}{\sqrt{2}} \cr} \right ),
\eea
(with $s_x \equiv \sin\theta_x$, $c_y \equiv \cos\theta_y$, and so
on) where the Euler rotation matrices are given by
\begin{eqnarray}
R_{12}(\theta_x) = \left ( \matrix{ c_x & s_x   & 0 \cr -s_x & c_x
& 0 \cr 0   & 0 & 1 \cr} \right ) \; &,&  R_{23}(\theta_y) = \left
( \matrix{ 1   & 0 & 0 \cr 0   & c_y & s_y \cr 0   & -s_y & c_y
\cr} \right ).
\end{eqnarray}
Substituting $V_\al$ with $V_0$ in (eq. \ref{star}), we get, in the flavor basis, a ``variant'' form of the
tripartite model:
\be
\label{gen_al}
  M_\nu =
  \pmatrix{ A_\al - B_\al + C_\al
  & {1\over {2\,\sqrt{2}}}\,\tan{(2\,\al)}\,C_\al & -{1\over {2\,\sqrt{2}}}\,\tan{(2\,\al)}\, C_\al \cr
  {1\over {2\,\sqrt{2}}}\,\tan{(2\,\al)}\,C_\al & A_\al + C_\al & B_\al - C_\al \cr
  - {1\over {2\,\sqrt{2}}}\,\tan{(2\,\al)}\,C_\al & B_\al - C_\al & A_\al + C_\al},
  \ee
  where
  \bea A_\al &=& -(3/4)\,\cos(2\,\al){(m_2 - m_1)}+(1/4)\,{(m_2 + m_1)}+(1/2)\,{ m_3}, \\
B_\al &=&
-(1/4)\,{ (m_2 + m_1)}+(3/4)\,\cos(2\,\al){ (m_2 - m_1)}+(1/2)\,{ m_3}, \\C_\al &=&\cos(2\,
\al){ (m_2 - m_1)}. \eea
We can check now that this variant form is completely determined by the invariance
under, say, $S_{3\al}$ and $S_{2\al}$ where
  \bea
\label{ZalZ2sy}
  S_{3\al} &\equiv& V_\al^* \cdot I_z \cdot V_\al^T = S_3,\\ 
  S_{2 \al} &\equiv& V_\al^* \cdot 
  I_y \cdot V_\al^T =
  \pmatrix{\cos{(2\,\al)} & -{1/\sqrt{2}}\, \sin{(2\,\al)}& {1/ \sqrt{2}}\, \sin{(2\,\al)} \cr
  -{1/ \sqrt{2}}\, \sin{(2\,\al)} & {1/ 2} -{1/ 2}\, \cos{(2\,\al)} & {1/ 2} +{1/ 2}\, \cos{(2\,\al)} \cr
  {1/ \sqrt{2}}\, \sin{(2\,\al)} & {1/ 2} + {1/ 2}\, \cos{(2\,\al)} & {1/ 2} - {1/ 2}\, \cos{(2\,\al)}}.
\eea

Sofar we have looked for symmetries which give results in line with the experimental data ($V_0$ in our case).
However, having
found a specific form of symmetry satisfying the experimental constraints we can now consider it as an ansatz 
for the underlying symmetry of the neutrino mass matrix and extend it, in a consistent manner,
to include other parts of the lepton sector.
In fact, the realization of the symmetry ($S\cong Z_2^3$) we have found on the left light
neutrinos should also apply to their doublet-charged lepton
partners, and it is often incompatible with a diagonal $M_l$ with
three different eigenvalues. In order to avoid this difficulty, we
introduce additional  scalars, which are singlet under the SM symmetry while transforming non-trivially under 
the proposed $S \cong Z_2^3$ symmetry.

\section{ The charged-lepton mass matrix}
We start with the normal SM mass term:
\bea \label{L1} {\cal{L}}_1 &=& Y_{ij} \overline{L}_i \Phi  l^c_j ,\eea
where the SM Higgs $\Phi$ and the charged right-handed leptons $l^c_j$ are assumed to be
singlets under the $S = \left(Z_2\right)^3$ symmetry, whereas the lepton
 left-doublets transform component-wise faithfully:
\begin{equation}
 L_i
\rightarrow S_{ij}L_j, 
\end{equation}
with $i,j =1,2,3$ and $S$ is the $\left(Z_2\right)^3$ symmetry given in 
eqs.~\ref{Z2cubesymmetry1}--\ref{Z2cubesymmetry3}.

The invariance under $S$ restricts the Yukawa-couplings to satisfy the matrix equation:
\bea \label{matrix equation} S^T\cdot Y&=& Y.\eea
This equation can not be met for a matrix $S$ with determinant equal to $-1$, and hence we deduce that the
$S\cong\left(Z_2\right)^3$-symmetry forces the term ${\cal{L}}_1$ to vanish, whereas it would have been 
allowed had we restricted the symmetry to $S_0\cong\left(Z_2\right)^2$.

In order to generate lepton masses, then,
we introduce three SM singlet scalar fields, $\Delta_k$, with nontrivial transformations under
the symmetry $S$, one for each family (the
indices $k=1,2,3$ refer also to the flavors $e$, $\mu$ and $\tau$ respectively).
The field $\Delta_k$ is coupled to the corresponding lepton left
doublet $L_k = \pmatrix{\nu_k \cr l_k}$ via the dimension 5 operator:
\begin{eqnarray}
{\cal{L}}_2 &=& \frac{f_{ikr}}{\Lambda} \overline{L}_i \Phi \Delta_k  l^c_r ,
\end{eqnarray}
where $\Lambda$ is a heavy mass scale. Note here that our {\it{ad hoc}} assumption of the coupling of charged
leptons with these additional Higgses via higher operators, and not through SM-like Yukawa terms, is apt
to reduce the effects of flavor changing neutral currents, usual when many Higgs doublets exist 
\cite{BjorWein77}. We assume the new scalars  $\Delta_k$ transforming under $S$ like the lepton left doublets:
\begin{equation}
 \Delta_i \rightarrow S_{i j}\Delta_{j}.
\end{equation}
Invariance of the Lagrangian under the symmetry implies
\begin{equation}
S_{i\alpha}\,S_{k\beta}\, f_{ikr} = f_{\alpha \beta r}.
\end{equation}
In matrix form, we write this as  
\bea
S^T \, f_r \, S =f_r , \eea
where $f_r$, for fixed $r$, is the matrix whose ($i,j$) entry is $f_{i j r}$. From (\ref{equivalence}), 
we have a solution for the new Yukawa coupling of the above equation in the form: 
\begin{eqnarray}
\label{cly} f_r=\left ( \matrix{ A_r - B_r + C_r  & C_r
& -C_r \cr C_r & A_r+C_r & B_r-C_r \cr -C_r & B_r-C_r &
A_r+C_r \cr} \right ) .
\end{eqnarray} 
When the fields $\Delta_k$ and $\phi^\circ$ take the vacuum expectation values (vevs)
$<\Delta_k >=\delta_k$ and $<\phi^\circ>=v$, the charged lepton mass
 matrix originating from ${\cal{L}}_2$ becomes:
\begin{equation}
\left( {\cal{M}}_2 \right)_{ir} = \frac{v f_{ikr}}{\Lambda}\delta_k.
\end{equation}
  As we are concentrating on the neutrino sector without stating explicitly the $\Delta_k$ potential and since
the $S$ symmetry is broken by ``soft'' terms in the Higgs sector, we may assume a 
$\Delta_3$-dominated pattern:
 $\delta_1,\delta_2 \ll \delta_3$, so to get the charged lepton mass matrix 
 \begin{eqnarray}\label{clm}
 M_l &=& \frac{v\delta_3}{\Lambda} \left ( \matrix{ -C_1  & -C_2 & -C_3 \cr B_1-C_1 &
B_2-C_2 & B_3-C_3 \cr A_1+C_1 & A_2+C_2 & A_3+C_3 \cr} \right ).
\end{eqnarray}
The determinant of $M_l$ is equal to:
$-\left(\frac{v\delta_3}{\Lambda}\right)^3 {\bf A} \cdot \left( {\bf B} \times {\bf C}\right)$,
where ${\bf A}$ is the vector of components $A_i$ (similarly  for ${\bf B,C}$), which means that 
these three vectors
should not be coplanar in order to have a nonsingular lepton mass matrix. We get then
\begin{eqnarray}\label{clmsquare}
M_l.M_l^\dagger &=&
\frac{v^2\delta_3^2}{\Lambda^2}{ \left ( \matrix{ {\bf C' . C'} & {\bf C' . B'} & {\bf C' .
A'} \cr {\bf B' . C'} & {\bf B' . B'} & {\bf B' . A'} \cr {\bf A'
. C'} & {\bf A' . B'} & {\bf A' . A'} \cr} \right )},
\end{eqnarray}
where ${\bf A'}= {\bf A} + {\bf C}$, ${\bf B'}= {\bf B} - {\bf C}$ and ${\bf C'}= -{\bf C}$.
If we just assume the magnitudes of the three vectors coming in ratios comparable to the lepton mass ratios:
\bea {\bf C'}^2:{\bf B'}^2:{\bf A'}^2 &\sim& m_e^2 : m_\mu^2 : m_\tau^2 \label{ratios},\eea then one can show
that the mixing $U^l_L$, such that $U^l_L M_l M_l^\dagger {U^l_L}^\dagger$ is diagonal, will be
naturally very close to the identity matrix with off-diagonal terms of order
($m_e/m_\mu \sim 5 \times 10^{-3}, m_e/m_\tau \sim 3 \times 10^{-4}, m_\mu/m_\tau \sim 6 \times 10^{-2}$), which
would mean that our basis is the flavor basis to a very good approximation and that the hierarchical charged
lepton masses can be obtained from a hierarchy on the a priori
arbitrary Yukawa couplings (${\bf C'}^2 \ll {\bf B'}^2 \ll {\bf
A'}^2$).

In order to clarify the last point, let us assume
\bea \frac{|{\bf C'}|}{|{\bf A'}|} = \lam_e &,& \frac{|{\bf B'}|}{|{\bf A'}|} = \lam_\mu,\eea
where $\lam_{e,\mu}$ are small parameters of order $m_{e,\mu}/m_\tau$. This yields the squared mass matrix
to be written as: \bea Q_\lam \equiv M_l M_l^\dagger &=& \frac{v^2 |{\bf A'}|^2 \delta_3^2}{\Lambda^2}
\left ( \matrix{
\lam_e^2 & \lam_e \lam_\mu \cos\psi & \lam_e \cos \phi \cr \lam_e \lam_\mu \cos\psi & \lam_\mu^2 &
\lam_\mu \cos \theta \cr \lam_e \cos \phi & \lam_\mu \cos \theta & 1 \cr} \right ),
\label{Q}
\eea
where $\theta$, $\phi$ are the angles between the vector ${\bf A'}$ and the vectors ${\bf B', C'}$ respectively, 
while $\psi$ is the angle between ${\bf B'}$ and ${\bf C'}$. The diagonalization of $M_l M_l^\dagger$ by
means of an infinitesimal rotation amounts to seeking an antisymmetric matrix
\be
I_\eps = \left ( \matrix{
0 & \eps_1 & \eps_2 \cr -\eps_1 & 0 & \eps_3 \cr -\eps_2 & -\eps_3 & 0 \cr} \right ),
\ee
with small parameters $\eps'$s,
satisfying: \bea \left( Q_\lam + \left[Q_\lam,I_\eps \right] \right))_{ij}&=0,& i\neq j.
\eea
Solving this equation analytically,
one can express the $\eps$'s in terms of ($\lam_{e,\mu},\cos (\psi,\phi,\theta)$). One finds, apart from
``fine tuned'' situations corresponding to coplanar vectors ${\bf A',B',C'}$, that we get:
$\eps_3 \sim \lam_\mu, \eps_2 \sim \lam_e$ and
$\eps_1 \sim \lam_e/\lam_\mu$, which indicates that a consistent solution for $U^l_L$ close to the identity 
matrix is
given by $U^l_L = e^{I_\eps} \approx I + I_\eps$. For example, taking the numerical values
$\lam_e = 3 \times 10^{-4}, \lam_\mu = 6 \times 10^{-2}$ and a common value $\pi/3$ for the angles
formed by the vectors, we get:
$m_e^2:m_\mu^2:m_\tau^2=6\times 10^{-8}:3\times 10^{-3}:1$, with the `exact' unitary diagonalizing matrix
given by:
\be
U^l_L \sim
\left ( \matrix{
1 & 10^{-3} & 10^{-4} \cr -1.6 \times 10^{-3} & 1 & 3 \times 10^{-2} \cr
 -10^{-4} & -3 \times 10^{-2} & 1 \cr} \right ).
 \ee
Thus, the deviations due to the rotations are, in general, small, but could justify measuring a nonzero small 
value of $U_{e3}$ which is restricted by the reactor data \cite{boehm} to be less than $0.16$ in magnitude.

\section{The neutrino mass matrix and type-I seesaw scenario }
In this scenario the effective light left neutrino mass matrix is generated through seesaw
mechanism as described in eq.~\ref{Majmass}.
The Dirac neutrino
mass matrix comes from the Yukawa term:
\begin{eqnarray} g_{ij} \overline{L}_i
\tilde{\Phi} \nu_{Rj},
\end{eqnarray}
where $\tilde{\Phi} = i \tau_2 \Phi^*$. As to the right neutrino, we will
assume that it transforms faithfully as
\begin{eqnarray} \label{rtransform}
\nu_{Rj} \rightarrow S_{j\gamma} \nu_{R\gamma},
\end{eqnarray} since, as we shall see,
this assumption will put constraints on the right Majorana mass
matrix. The invariance of the Lagrangian under $S$ implies in matrix form:
\begin{eqnarray}
S^T. g.S=g .
\end{eqnarray}
Thus, the relation (\ref{equivalence}), with $S$ standing for the $S_i$'s, enforces the Yukawa couplings
$g$ to be symmetric and the neutrino Dirac mass matrix
to have the form:
\begin{eqnarray}
\label{nDm}
M_\nu^D &=& v  \left ( \matrix{ A_D - B_D + C_D  &
C_D & -C_D \cr C_D & A_D+C_D & B_D-C_D \cr -C_D & B_D-C_D &
A_D+C_D \cr} \right ).
\end{eqnarray}

As to the
right-handed Majorana mass matrix, it comes from the term:
\begin{eqnarray}
\frac{1}{2} \nu^T_{iR} C \left( M_R \right)_{ij} \nu_{jR},
\end{eqnarray}
where
$C$ is the charge conjugation matrix. Again, the invariance under
the transformation (\ref{rtransform}) implies
\begin{eqnarray}
S^T\cdot M_R \cdot S &=&
M_R,
\end{eqnarray}
and thus $M_R$, which has to be symmetric, has the form:
\begin{eqnarray}
\label{nRm}
M_R &=& \Lambda_R \left ( \matrix{ A_R - B_R + C_R  &
C_R & -C_R \cr C_R & A_R+C_R & B_R-C_R \cr -C_R & B_R-C_R &
A_R+C_R
 \cr} \right ).
 \end{eqnarray}

Using equations (\ref{Majmass},\ref{nDm},\ref{nRm}), we have the
effective neutrino mass matrix:
\begin{eqnarray}\label{nm} M_{\nu} &=&
-\frac{v^2}{\Lambda_R} \left ( \matrix{ A_{\nu} - B_{\nu} + C_{\nu}  & C_{\nu} &
-C_{\nu} \cr C_{\nu} & A_{\nu}+C_{\nu} & B_{\nu}-C_{\nu} \cr -C_{\nu} & B_{\nu}-C_{\nu} &
A_{\nu}+C_{\nu}
 \cr} \right ),
 \end{eqnarray}
 where
 \begin{eqnarray} \label{coef} A_{\nu} &=& 
 \frac{-2A_DB_RB_D+A_RA_D^2+A_RB_D^2}{\left(A_R-B_R\right)\left(A_R+B_R\right)}, \nonumber \\
 B_{\nu} &=& \frac{-B_D^2B_R-A_D^2B_R+2A_RA_DB_D}{\left(A_R-B_R\right)\left(A_R+B_R\right)}, \\
 C_{\nu} &=& \frac{-C_R\left(A_D-B_D\right)^2+C_D\left[3C_D+2\left( A_D-B_D \right)\right]\left(A_R-B_R\right)}
 {\left(A_R-B_R\right)\left(A_R-B_R+3C_R\right)}. \nonumber
 \end{eqnarray}
The neutrino mass eigenvalues are given by equation (\ref{neigenm}):
\begin{equation}
 \frac{v^2}{\Lambda_R}\left(A_{\nu}-B_{\nu},\; A_{\nu}-B_{\nu}+3C_{\nu},\; A_{\nu}+B_{\nu}
 \right).\end{equation}

All different patterns of the neutrino masses can accommodated, as follows:
\begin{itemize}
\item{\it Normal hierarchy}: It suffices to have
\begin{eqnarray} A_i \simeq
B_i,\; C_i \ll B_i,\; i=R,D,
\end{eqnarray}
for getting a normal hierarchy with
\begin{eqnarray}
A_{\nu} \simeq \frac{A_D^2}{A_R},\; B_{\nu} \simeq \frac{B_D^2}{B_R},\; C_{\nu}
\simeq \frac{C_D^2}{C_R}.
\end{eqnarray}
We see that one can arrange the
Yukawa couplings to enforce $A_{\nu} \simeq B_{\nu},\;C_{\nu} \ll B_{\nu}.$

\item{\it Inverted hierarchy}: It is sufficient to have
\begin{eqnarray} A_i
\simeq -B_i,\; C_i \ll B_i,\; i=R,D,
\end{eqnarray}
so that one gets an inverted
hierarchy with
\begin{eqnarray}
A_{\nu} \simeq \frac{A_D^2}{A_R},\; B_{\nu} \simeq
\frac{B_D^2}{B_R},\; C_{\nu} \simeq \frac{2C_DA_DA_R-C_RA_D^2}{A_R^2}.
\end{eqnarray}
One can arrange the Yukawa couplings to enforce $A_{\nu} \simeq
-B_{\nu},\; C_{\nu} \ll B_{\nu}.$

\item{\it Degenerate case}: If we have
\begin{eqnarray}
A_i \gg B_i \gg C_i,\;
i=R,D,\end{eqnarray}
then we get
\begin{eqnarray} A_{\nu} \simeq \frac{A_D^2}{A_R},\; B_{\nu} \simeq
\frac{2 A_D B_D}{A_R}-\frac{A_D^2 B_R}{A_R^2},\; C_{\nu} \simeq
\frac{2A_DA_RC_D-C_RA_D^2}{A_R^2}.
\end{eqnarray}
One can arrange the Yukawa
couplings to enforce $A_{\nu} \gg B_{\nu} \gg C_{\nu}$, so that we have a
degenerate spectrum.
\end{itemize}
\vspace{-3mm}
Thus, we see that a certain pattern occurring in both the Dirac and
the right-handed Majorana mass matrices can resurface in the
effective neutrino mass matrix.

The right handed (RH) neutrino mass term violates lepton number by two units. The out of equilibrium decay of 
the lightest RH neutrino  to standard model particles can be a natural source of lepton asymmetry \cite{FY} 
and it is given by
\begin{eqnarray}
\epsilon \simeq \frac{3}{16\pi
v^2}\frac{1}{(M_\nu^{D\dagger}M_\nu^D)_{11}}\sum_{j=2,3}
\mbox{Im}[\{(M_\nu^{D\dagger}M_\nu^D)_{1j}\}^2]\frac{M_{R1}}{M_{Rj}},
\end{eqnarray}
where $M_{Ri},\; i=1\cdots 3$ are the masses for right handed neutrinos.
Explicitly we have
\begin{eqnarray}
(M_\nu^{D\dagger}M_\nu^D)_{12} &=& 3|C_D|^2 + (C_DA^*_D + C^*_DA_D) - (C_DB^*_D + C^*_DB_D),\nonumber \\
(M_\nu^{D\dagger}M_\nu^D)_{13} &=& -3|C_D|^2 - (C_DA^*_D + C^*_DA_D) + (C_DB^*_D + C^*_DB_D),
\end{eqnarray}
 which gives a vanishing  lepton asymmetry . Thus, in this seesaw type mechanism the baryon asymmetry 
 is zero if $\left(Z_2 \right)^3$ is an exact symmetry.

\section{The neutrino mass matrix and type-II seesaw scenario}
In this scenario we introduce two SM triplet fields $\Sigma_A$, $A=1,2$ which are also assumed to be singlet
under the flavor symmetry $\left(Z_2 \right)^3$. The Lagrangian part relevant for the neutrino mass matrix is
\begin{equation}
{\cal{L} } = \lambda_{\alpha\beta}^{A}\, L_\alpha^T\, C\, \Sigma_A\, i\,\tau_2\, L_\beta +
{\cal{L}}(H,\Sigma_A)+h.c.,
\end{equation}
where $A=1,2$ and
\begin{eqnarray}
{\cal{L}}(H,\Sigma_A) &=& \mu_H^2 H^\dagger H + \frac{\lambda_H}{2} {(H^\dagger H)}^2+
M_A\, \mbox{Tr}\left(\Sigma_A^\dagger \Sigma_A \right)+
\frac{\lambda_{\Sigma_A}}{2} \left[\mbox{Tr}\left( \Sigma_A^\dagger \Sigma_A\right)\right]^2
+ \\\nonumber && \lambda_{H\Sigma_A} (H^\dagger H) \mbox{Tr}\left( \Sigma^\dagger_A \Sigma_A\right)
+
{\mu_A H^T \Sigma_A^\dagger i\tau_2 H +h.c.},
\end{eqnarray}
where $H=\pmatrix{\phi^+ \cr \phi^0}$, and
\begin{eqnarray}
\Sigma_A &=& \left ( \matrix{ \frac{\Sigma^+}{\sqrt{2}}  &
\Sigma^0 \cr \Sigma^{++} & -\frac{\Sigma^+}{\sqrt{2}}\cr} \right )_A.
\end{eqnarray}

The neutrino mass matrix due to the exchange of the two triplets, $\Sigma_1$ and $\Sigma_2$, is
\begin{equation} \label{mass}
(M_\nu)^A_{\alpha\beta}\simeq v^2 \left[\lambda^1_{\alpha\beta} \frac{\mu_1}{M^2_{\Sigma_1}} +
\lambda^2_{\alpha\beta}\frac{\mu_2}{M^2_{\Sigma_2}}\right],
\end{equation}
where $M_{\Sigma_i}$ is the mass of the neutral component $\Sigma_i^0$ of the triplet $\Sigma_i ,i=1,2$.

Here some remarks are in order. First,  the symmetry $Z_2^3$ implies that $\lambda_1$ and $\lambda_2$ 
have the same tripartite structure. Second, due to the `tadpole' term (the $\mu_A$-term) 
in ${\cal{L}}(H,\Sigma_A)$, which would forbid the ``unwanted'' spontaneous breaking of the lepton number, 
one can arrange the parameters so that minimizing the potential gives a non-zero vev for the neutral 
component $\Sigma^0$ of the triplet. This would generate a mass term for the neutrinos, however, 
the procedure here is equivalent to integrating out the the heavy triplets and both ways lead to
the mass formula above. Third, the flavor changing  neutral current due to the triplet is highly
suppressed due to the heaviness of the triplet mass scale, or equivalently the smallness of the 
neutrino masses.

Now let us discuss the baryon asymmetry. We will show that even though the neutrino Yukawa couplings are real 
it is possible to generate a  baryon to photon density consistent with the observations. Since the triplet 
$\Sigma_A$ can decay into lepton pairs $L_\alpha L_\beta$ and $HH$, it implies that these processes violate 
total lepton numbers (by two units) and may establish a lepton asymmetry. As the universe cools further, 
the sphaleron interaction \cite{KRS} converts this asymmetry into baryon asymmetry. At temperature of the 
order $\mbox{max}\{M_1, M_2\}$, the heaviest triplet would decay via  lepton number  violating interactions. 
However, no asymmetry will be generated from this decay since the rapid  lepton number violating interactions 
due to the lightest Higgs triplet will erase any previously generated  lepton asymmetry. Thus, only when the 
temperature becomes just below the mass of the lightest triplet Higgs the asymmetry would be generated.

With just one triplet, the lepton
asymmetry will be generated at the two loop level and it is highly suppressed. The reason is that one can 
always redefine the phase of the Higgs field to render the $\mu$ real which will result in the vanishing 
of the absorptive part of the self energy diagram. The choice of having more than one Higgs triplet is 
necessary to generate the asymmetry \cite{HMS}. In this case,  the CP asymmetry in the decay of the lightest 
Higgs triplet (which we choose to be $\Sigma_1$) is generated at one loop level due to the interference 
between the tree and the one loop self energy diagram \footnote { There is no one loop vertex correction 
because the triplet Higgs is not self conjugate}  and it is given by
\begin{equation}
\epsilon_{CP} \approx -\frac{1}{8\pi^2} \frac{\mbox{Im}\left[\mu_1\mu_2^*
\mbox{Tr}\left(\lambda^1\lambda^{2\dagger}\right)\right]}{M_2^2}
\frac{M_1}{\Gamma_1},
\end{equation}
where $\Gamma_1$ is the decay rate of the lightest Higgs triplet and it is given by
\begin{eqnarray}
\Gamma_1 = \frac{M_1}{8\pi}\left[\mbox{Tr}\left(\lambda^{1\dagger} \lambda^1\right) 
+ \frac{\mu_1^2}{M_1^2}   \right].
\end{eqnarray}
The baryon to photon density is approximately given by
\begin{equation}
\eta_B \equiv \frac{n_B}{s} =\frac{1}{3}\eta_L \simeq \frac{1}{3} \frac{1}{g_*K} \epsilon_{CP},
\end{equation}
where $g_* \sim 100$ is the number of relativistic degrees of freedom at the time when the Higgs triplet
decouples from the thermal bath and $K$ is the efficiency factor \cite{Buchmuller} defined as
\begin{eqnarray}
K =\frac{\Gamma_1}{H}(T= M_1),
\end{eqnarray}
 ($H$ is the Hubble parameter) which takes into account the fraction of out-of equilibrium decays 
 and the washout effect. For $\mu_1 \approx M_{\Sigma_1} \sim 10^{12}\, \mbox{GeV}$ the efficiency 
 factor is of order $10^{-3}$  and the baryon asymmetry is
\begin{equation}
\eta_B\approx  10^{-7}\frac{
\mbox{Tr}\left(\lambda^1\lambda^{2\dagger}\right)}{\mbox{Tr}\left(\lambda^{1\dagger} \lambda^1 \right) + 1}
\sin(\phi_2 -\phi_1).
\end{equation}
Thus one can produce the correct baryon-to -photon ratio of $\eta_B \simeq 10^{-10}$ by choosing 
$\lambda$'s of order $0.1$ and not too small relative phase between the $\mu$'s.

\section{Summary}
We showed in a simple way that the underlying symmetry of the tripartite model is
$Z_2^3$ and we implemented this symmetry in a complete
setup including the charged leptons, the neutrinos, and extra scalar fields. We
showed that this setup can accommodate the different patterns of
charged leptons and neutrino mass matrices. We showed that one can produce the correct baryon-to -photon ratio 
in type-II seesaw mechanism by choosing appropriate couplings and not too small relative phase between the 
vacuum expectation values of the extra scalar fields.

\section*{Acknowledgements}
Major parts of this work were done within the Associate Scheme of
ICTP. We thank E. Ma, S. Petcov and A. Smirnov for useful discussions. N. C. thanks
CBPF-Brazil, where part of the work has been done, for its hospitality and acknowledges support from TWAS.

\bibliographystyle{mdpi}

\end{document}